\documentclass[lettersize,journal]{IEEEtran}
\IEEEoverridecommandlockouts
\usepackage{acronym}
\usepackage{amsmath,amssymb,amsfonts}
\usepackage{booktabs}
\usepackage{cite}
\usepackage{graphicx}
\usepackage{subcaption}
\usepackage{tabularx} 
\usepackage{soul}
\usepackage{textcomp}
\usepackage{xspace}
\usepackage{algpseudocode}
\usepackage[linesnumbered,ruled,vlined]{algorithm2e}
\usepackage{noconflict}
\usepackage{mathtools}
\usepackage{url}
\usepackage{array}
\usepackage{stfloats}
\usepackage{verbatim}
\usepackage{enumitem}
\usepackage{booktabs} 
\usepackage{makecell} 
\usepackage[table]{xcolor} 
\usepackage{multirow}
\usepackage{orcidlink}
\usepackage{hyperref}
\usepackage{academicons}
\usepackage{xcolor}
\begin{document}

\title{Analysis of DNS Dependencies and their Security Implications in Australia: A Comparative Study of General and Indigenous Populations}

\author{Niousha~Nazemi~\orcidlink{0000-0003-4085-7044},~\IEEEmembership{Member,~IEEE,}
        Omid~Tavallaie~\orcidlink{0000-0002-3367-1236}, 
        Anna~Maria~Mandalari~\orcidlink{0000-0002-6715-101X},
        Hamed~Haddadi~\orcidlink{0000-0002-5895-8903},~\IEEEmembership{Member,~IEEE,}
        Ralph~Holz~\orcidlink{0000-0001-9614-2377},
        and~Albert~Y.~Zomaya~\orcidlink{0000-0002-3090-1059},~\IEEEmembership{Fellow,~IEEE}}

\maketitle

\newcommand{\eg}[0]{e.g.,\xspace}
\newcommand{\etc}[0]{etc\xspace}
\newcommand{\todo}[2]{\hl{TODO (#1): #2}}

\begin{abstract}
This paper investigates the impact of internet centralization on DNS provisioning, particularly its effects on vulnerable populations such as the indigenous people of Australia. We analyze the DNS dependencies of Australian government domains that serve indigenous communities compared to those serving the general population. Our study categorizes DNS providers into leading (hyperscaler, US-headquartered companies), non-leading (smaller Australian-headquartered or non-Australian companies), and Australian government-hosted providers. Then, we build dependency graphs to demonstrate the direct dependency between Australian government domains and their DNS providers and the indirect dependency involving further layers of providers. Additionally, we conduct an IP location analysis of DNS providers to map out the geographical distribution of DNS servers, revealing the extent of centralization on DNS services within or outside of Australia. Finally, we introduce an attacker model to categorize potential cyber attackers based on their intentions and resources. By considering attacker models and DNS dependency results, we discuss the security vulnerability of each population group against any group of attackers and analyze whether the current setup of the DNS services of Australian government services contributes to a digital divide.
\end{abstract}


\section{Introduction}
\label{sec: Introduction}

\IEEEPARstart{T}{he} concept of Internet centralization \cite{arkko2019considerations} refers to consolidating control over the Internet's infrastructure and services. The Domain Name System (DNS) is a core Internet service. In the context of the DNS, centralization refers to the control that is held by a limited number of dominant companies \cite{emergen_2023_dns} such as Amazon Web Services (AWS) Route 53 \cite{AmazonRoute53}, Microsoft Azure \cite{AzureDNS}, Cloudflare DNS\cite{CloudflareDNS}, Google Cloud DNS \cite{GoogleCloudDNS}, and Akamai Edge DNS \cite{AkamaiEdgeDNS}. This centralization leads to secondary effects on Internet services and infrastructure, such as the availability of the daily services used by a large population of Internet users \cite{rfc9518}.  

In this regard, the significance of authoritative name servers becomes apparent in the context of service dependencies. The latter term refers to a situation where a service's functionality or effectiveness relies on another's correct functioning. Hence, the efficiency and accuracy of the DNS resolution process are highly dependent on the performance and security of the authoritative name servers. A single point of failure is a major risk in such a system. For example, if an authoritative name server in a centralized setting fails, it can disrupt access to many websites and online services despite redundancies. This is not impossible, and scenarios with cascading failures due to centralization have been observed in the recent past~\cite{crowdstrike2024,facebook2021}.

In this paper, we ask how failures due to DNS centralization can impact different population groups. We choose the Australian setting as our focus area for two reasons: first, Australia has a sizable vulnerable population, namely the indigenous community. Second, Australia has a highly developed digital infrastructure, and many government services are available online. We analyze the DNS dependencies of Australian government domains that serve Indigenous communities compared to those serving the general population. Our study makes three main contributions:
\begin{itemize} [left=0pt]

\item \textbf{Digital Divide:} We examine whether there is a difference in the DNS service providers used for indigenous and general population domains. Our purpose is to understand whether such variations contribute to a possible digital divide between the general population and indigenous peoples.

\item \textbf{Geographical Dependencies:}
We analyze the geographical distribution of DNS servers linked to Australian governmental domains. We focus particularly on servers operated by hyperscaler companies and aim to determine these servers' physical locations, distinguishing those within Australia from those abroad.

\item \textbf{Security Problems:} By considering the DNS dependency results and the provider IP networks, we discuss the vulnerability of each domain group to different types of attacker models. We categorize potential attackers into representative groups of varying strengths and analyze their potential impact on different domain groups.

\end{itemize}

\begin{figure}[t]
  \centering
  \begin{minipage}[b]{0.30\linewidth}
    \includegraphics[width=\linewidth]{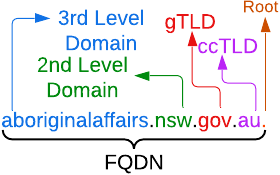}
    \caption{FQDN.}
    \label{fig:FQDN}\vspace{-2mm}
  \end{minipage}
  \hfill
  \begin{minipage}[b]{0.65\linewidth}
    \includegraphics[width=\linewidth]{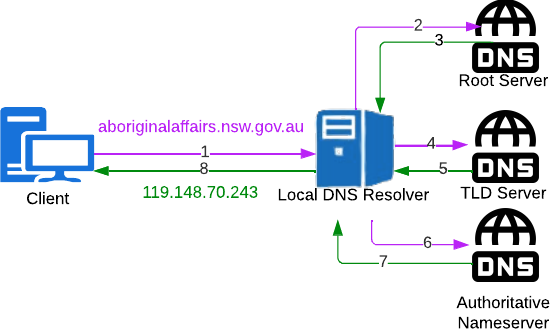}
    \caption{DNS resolution process.}
    \label{fig:DNS_Resolution}\vspace{-2mm}
  \end{minipage}
\end{figure}

Our methodology consists of four main phases: Australian government service identification, data collection, retrieving authoritative name servers, and constructing DNS dependency graphs. We categorize DNS providers into leading (hyperscaler, US-headquartered companies), non-leading (smaller Australian-headquartered or non-Australian companies), and Australian government-hosted providers. This paper is an extended version of our previous work \cite{nazemi2023dns} presented at the Workshop on Transparency, Accountability, and User Control for a Responsible Internet (TAURIN) in 2023 \cite{taurin2023}, where focused exclusively on the digital divide, we now broaden our scope significantly to include security analyses and analysis of geographic dependencies.

The paper is structured as follows: We begin with background information on DNS and related concepts, followed by a review of related work. We then detail our methodology and present our results, focusing on direct dependency analysis, indirect dependency analysis, and geographic analysis. The discussion section interprets these results in the context of the digital divide, geographical dependencies, and security implications, including an analysis of attacker models. We conclude with a summary of our findings and their significance for DNS provisioning in Australia.

\section{Background}

We provide background for our study of DNS dependencies and their impact on Australian government domains, particularly those serving indigenous communities.

\subsection{DNS and Its Hierarchy}
DNS \cite{bib13_RFC1034, bib40_RFC1035} is a fundamental Internet component responsible (among other things) for translating human-readable domain names into machine-readable IP addresses. Dependencies expressed in the DNS are critical for the availability of Internet services.
The DNS hierarchy begins with root servers \cite{IANA}, followed by Top-Level Domains (TLDs), which include generic TLDs (gTLDs) like .com and country-code TLDs (ccTLDs) such as .au for Australia. Below these are second-level domains. The various parts of a Fully Qualified Domain Name (FQDN) are given as the concatenation of so-called labels at the various levels of this hierarchy. Fig. \ref{fig:FQDN} shows an example.
The DNS resolution process, illustrated in Fig. \ref{fig:DNS_Resolution}, involves multiple steps. In the simplest case, and without caching, this includes: 1) A recursive DNS server initiates the query. 2) It queries a root server, which then directs it to the corresponding TLD server. 3) The TLD server directs the query to the authoritative name server (NS) for the specific queried domain. 4) The authoritative server retrieves the IP address, which is subsequently sent back to the recursive server. User interaction happens via a stub DNS resolver that directs its queries to the recursive server, which is usually provided by an ISP or operated as a public service (e.g., the public resolver of Google).

\subsection{DNS Dependencies}
\begin{figure}[t]
\centering
  \mathtoolsset{showonlyrefs=false}  \includegraphics[width=\columnwidth]{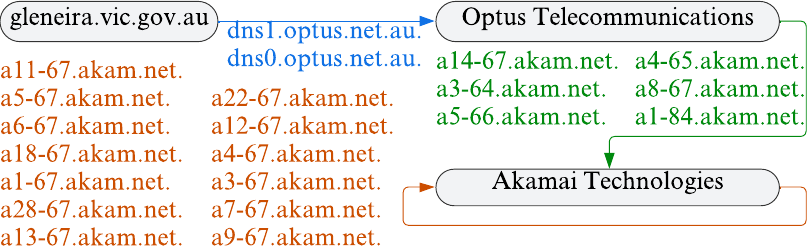}
  \caption{DNS Dependencies: lines represent name servers.}
  \label{fig:dgraph_gen_exp}\vspace{-1 mm}
\end{figure}

Service dependencies \cite{bib12_kashaf2020analyzing, bib10_ikram2019chain} can be categorized into direct and indirect forms as follows.
\textbf{Direct dependencies} occur when a service (e.g., a website) explicitly relies on other services provided by a third-party company. For the DNS, an example would be website $X$ designating company $Y$ as its DNS provider, which results in $Y$ operating the authoritative name servers for $X$'s domain.
\textbf{Indirect dependencies} arise from a chain (or multiple chains) of direct dependencies. If site $X$ uses provider $Y$, which in turn depends on company $Z$, then $X$ indirectly depends on $Z$. As it is possible to have authoritative name servers with multiple companies, many such chains can exist. Consequently, a service may be unknowingly exposed to vulnerabilities in these chains. Fig. \ref{fig:dgraph_gen_exp} illustrates these dependencies using the example of \textit{gleneira.vic.gov.au}. Despite not directly using Akamai's services, this domain could be affected by Akamai outages due to its indirect dependency through Optus \cite{optus}.

\subsection{Digital Divide and Indigenous Australians}

The term digital divide refers to the gap between individuals or groups who have access to and can effectively use digital services and those who lack such access \cite{van2020digital}. It affects economically disadvantaged and marginalized groups, including the indigenous people of Australia \cite{samaras2005Indigenous}. Indigenous Australians, comprising Aboriginal peoples and Torres Strait Islander peoples \cite{firstnationsvocabulary}, often face challenges in accessing quality Internet services due to their remote locations~\cite{singleton2009youth}. While this client-side problem is well known, the geographical isolation can be exacerbated by inadequate DNS infrastructure on the service side. This may potentially lead to a different form of digital divide, which has so far received very little attention. By examining the DNS infrastructure serving the indigenous communities, we can identify potential disparities in service quality, reliability, and security that may arise from different dependency patterns or geographical distributions of DNS servers. We note that the indigenous people of Australia are also referred to as \textbf{Indigenous Australians} or the \textbf{First Nations people} \cite{firstnationsvocabulary}. In this paper, we use the term \textbf{indigenous}, which is also used by the United Nations \cite{UNDRIP}.

\section{Related Work} \label{sec: Related Work}
Many studies have explored the digital divide such as \cite{wang2022impact, singleton2009youth, intahchomphoo2018Indigenous}. Wang et al. \cite{wang2022impact} examined the digital divide in the context of energy poverty, revealing its negative effects on Internet utility. In Australia, the digital divide has emerged due to the significant remoteness and isolation of indigenous communities, which reduces the quality of Internet connectivity and limits their access to digital services \cite{singleton2009youth, intahchomphoo2018Indigenous}.

The phenomenon of centralization on the Internet and the concentration of infrastructure in the hands of a few dominant service providers have been the subject of investigation in previous studies. Zemburki et al. in \cite{zembruzki2022hosting} analyzed market share and top DNS providers. The authors used active measurement to examine this trend across different TLDs and geographical locations by using datasets obtained from the OpenINTEL \cite{OpenINTEL} project. Their results revealed a growing concentration of Internet infrastructure over time. In \cite{zembruzki2022consolidation}, Zembruzki et al. conducted an analysis of DNS consolidation by evaluating authoritative name server configurations across multiple ccTLDs and gTLDs which utilized five years from the OpenINTEL project. According to their findings, the top 5 DNS providers handle 20\% of all analyzed domain names, while the top 100 providers manage 80\% of domains with resolvable IPv4 addresses.
The authors of \cite{bib28_moura2020clouding} analyzed the DNS traffic from the B-Root DNS root server and two ccTLDs, namely New Zealand's .nz and the Netherlands' .nl. They measured the level of centralization resulting from the dominant cloud providers, including Amazon, Google, Microsoft, Cloudflare, and Facebook. 
Their study revealed that over one-third of queries made to both ccTLDs 
are the IP addresses of these providers.

The authors of \cite{bib28_moura2020clouding, bib12_kashaf2020analyzing} also studied the concept of dependencies and their impact on the Internet ecosystem. In \cite{bib12_kashaf2020analyzing}, the authors analyzed direct dependencies by focusing on the Mirai Dyn attack and its impact on the top 100k Alexa websites \cite{alexa} that relied on the Dyn DNS server as their DNS provider. Their findings revealed that about 90\% of the top 100K websites listed by Alexa depend on third parties. Among these websites, 50\% to 70\% are prone to service outages if prominent DNS providers (Cloudflare, AWS DNS, and DNSMadeEasy) fail. In \cite{bib42_deccio2010measuring}, \cite{bib43_deccio2012quantifying}, and \cite{bib44_deccio2009quality}, the authors developed a graph-based model to investigate the phenomenon of dependencies within the DNS context. Their findings demonstrated that over 50\% of domain name resolution processes rely on third-party authoritative name servers. Xu et al. in \cite{bib45_xu2022name} developed a graph model to illustrate the dependency relationship between domains and servers in the name resolution process. Additionally, they introduced a tool to determine the smallest set of name servers whose failure could disrupt resolution for each domain. Our study follows a similar approach, but due to the much smaller sample size, we do not build a dedicated graph model.

Furthermore, studies such as \cite{kashaf2023first} and \cite{bib16_urban2020beyond} have investigated the impact of third-party dependencies on services including DNS, CDNs (Content Delivery Networks), and CAs (Certificate Authority). Kashaf et al. in \cite{kashaf2023first} focused on African-centric websites. They used a combination of web crawling, data analysis, and visualization techniques. Their results indicated that African websites are significantly affected by third-party dependencies, with a significant concentration in reliance on third-party service providers. The authors of \cite{bib16_urban2020beyond} investigated website dependencies more precisely. Similar to \cite{bib10_ikram2019chain}, this work developed the concept of a chain of dependencies called TPTs (Third Party Trees) to reveal the hierarchical dependencies between each website and its third-party service providers.
Moura et al., in their study \cite{bib29_moura2021tsuname}, identified a vulnerability called TsuNAME in the DNS system. This vulnerability leads to loops in the resolution process and creates cyclic dependencies that generate a high frequency of queries. The authors demonstrated how such query flooding significantly increases traffic forwarding to authoritative servers---this escalation in traffic results in potential denial of service to DNS services. By studying real-world events within New Zealand's .nz ccTLD, the authors found that the misconfiguration of just two domains leads to a  50\% surge in overall traffic.

In contrast to these works, our paper aims to investigate the implications of DNS dependencies on indigenous populations in Australia, with a particular focus on the cascading effects that arise from relying on DNS hosting providers. Our analysis is centered around services delivered to people through government websites. To the best of our knowledge, this research is the first of its kind to explore the impact of DNS dependencies through the lens of services to vulnerable indigenous communities in Australia.

\newcolumntype{Y}{>{\centering\arraybackslash}X}

\begin{table*}
\caption{Keywords sorted based on the identified 16 categories of government services.}\label{tab:keywords}
\begin{tabularx}{\linewidth}{|Y|Y|Y|Y|}
\hline
\rowcolor{gray!25} 
\textbf{Healthcare} & \textbf{Disability Support} & \textbf{Family Support} & \textbf{Business Support}\\
\hline
Preventive care & Rehabilitation services & Child support & Business training \\
Chronic conditions & Assistive technology & Residential care & Business mentoring\\
Specialist care & Improve accessibility & Childcare & Entrepreneurship \\
Telehealth services & Promote social inclusion  & Youth support & Procurement policies \\
Vaccination & Community program  & Adolescent support & Provide funding\\
Medical services & & Violence prevention & \\
\hline
\rowcolor{gray!25}
\textbf{Education} & \textbf{Housing} & \textbf{Community Development} & \textbf{Environmental Programs} \\
\hline
Training programs & Homelessness support & Individual support & Land management\\
School programs & Affordable housing & Cultural maintenance  & Protect sacred sites\\
Vocational training & Appropriate housing & Social connection & Traditional lands \\
Adult education & Home-ownership & & Natural resources \\
 & & & Preserve cultural heritage\\
\hline
\rowcolor{gray!25}
\textbf{Disaster Relief} & \textbf{Economic Development} & \textbf{Women Support} & \textbf{Legal Services}\\
\hline
Emergency services & Employment services & Women health & Legal aid \\
Rebuilding homes & Job training & Accommodation service & Resolving disputes \\
Infrastructure improvement & Job seeking & Support groups & Justice \\
Temporary accommodation & Financial assistance & Employment opportunities & Family law \\
Distribution of food & Financial stability & Domestic violence & Criminal law\\
\hline
\rowcolor{gray!25}
\textbf{Retirement Support} & \textbf{Cultural Preservation} & \textbf{Mental Health} & \textbf{Addiction Support} \\
\hline
Age pension & Language program & Well-being & Substance abuse \\
Superannuation savings & Traditional arts and crafts & Counselling services & Treatment service\\
\hline
\end{tabularx}
\end{table*}
 
\section{Methodology}
We present the methodology employed in our paper to investigate the impact of DNS dependencies on websites serving the indigenous people of Australia. Our methodology consists of four main phases: Australian government service identification, data Collection, retrieving authoritative name servers, and constructing DNS dependency graphs.

\subsection{Service Category Exploration}
\label{sec: Service Category Exploration}

The initial phase involves desk research on online resources (such as government web pages, news websites, and blog posts) to identify the scope of services the Australian government offers to both general and indigenous populations. Our exploration begins with \textit{Services Australia}\cite{ServicesAustralia} portal, a foundational government resource that includes information about a considerable range of services accessible to Australians. Despite the information provided by \textit{Services Australia} and other government resources, to the best of our knowledge, there is no single, comprehensive, and categorized list of services that could be directly utilized for our research. Therefore, we conducted desk research to identify the \textbf{broad categories of government services} relevant to our study. This preliminary exploration lays the groundwork for the subsequent data collection phase. We identified 16 main categories of government services. These categories and their corresponding keywords are presented in Table~\ref{tab:keywords}. Each category is extracted by initial analysis of the web page's content to identify the most relevant and frequently occurring terms and phrases that best describe the services within each category. These categories include a wide range of essential services. For instance, Australia's \textbf{healthcare} system operates as a complex network involving public and private sector organizations that work together to provide health services to all Australians \cite{aihw_website}. In parallel to healthcare, the Australian government offers \textbf{support services for individuals with disabilities}, ranging from specialized medical care to providing assistive technology and financial assistance to enhance the quality of life and ensure inclusion in society \cite{dss_disability_services}.
These selected categories serve as the primary keywords for our data collection process to identify the specific government domains that deliver essential services to the general and indigenous populations in Australia. By focusing on these services and finding the relevant government domains, particularly in relation to the indigenous population, we understand how DNS dependencies and geographic distribution of servers might affect access to essential government services.

\subsection{Domain Collection} \label{sec: Data Collection}
In this section, we outline the domain collection process, specifically targeting Australian government websites that provide a range of services to the general and indigenous population. The types of services identified in section \ref{sec: Service Category Exploration} serve as the primary keywords for this process. By utilizing these principal keywords, we identify the actual government services and their corresponding domains. There are no open-access data sets of Australian domains categorized based on the target population group that we could have used directly for our study. Hence, we employ a web crawling technique to identify them.  We follow these steps to collect governmental domains for the general and indigenous people of Australia: 

\noindent 1) Web search: Utilizing Google search, we fetch pertinent governmental websites using keywords associated with the 16 service categories identified in the previous step. These keywords, derived from the service category exploration phase, serve as search queries to identify government websites offering services in these specific areas. Our search is restricted to websites with the \textit{.gov.au} suffix, guaranteeing the inclusion of only authorized Australian government websites. 

\noindent 2) URL extraction: We identify the first 100 URLs obtained from Google search results and add them to the URL dataset. 

\noindent 3) Keyword extraction: This step involves crawling the URLs of the visited web pages and using a word cloud technique to extract the top five most common and contextually relevant words from each web page. A manual check is then conducted to verify relevancy. These extracted keywords are cross-checked against the existing ones in the set, and those that are new and relevant are added to the set for further searches. 

\noindent 4) Domain extraction: we parse the web page URLs and extract the corresponding domains, which are then collected and stored in a dedicated data set. 

\noindent 5) Iteration: We iterate until no additional keywords or domains are identified. 

\begin{figure*}[t]
\centering  
    \includegraphics[width=\textwidth]{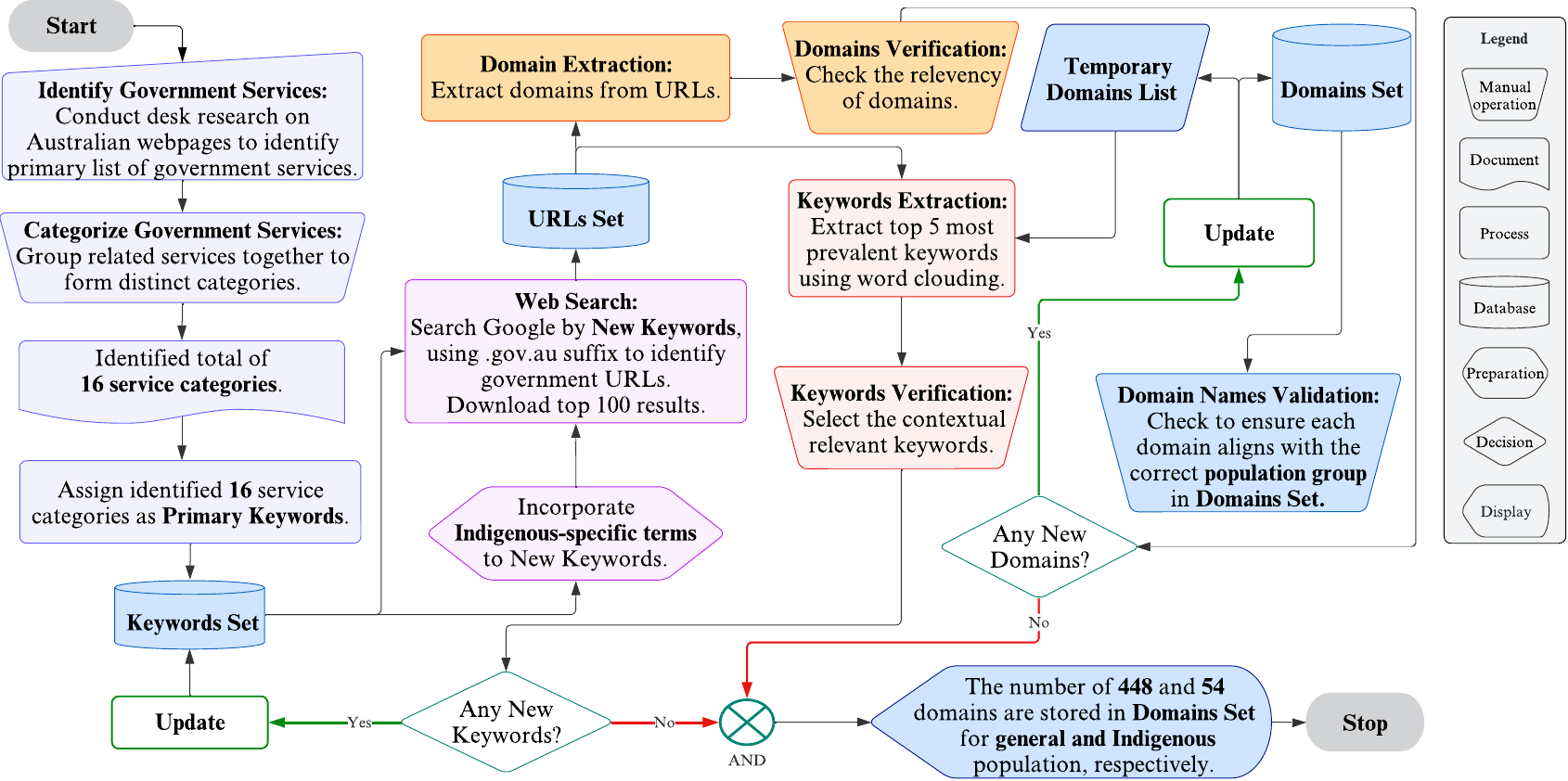}    
    \caption{Flowchart for identifying and categorizing Australian government domains by general and indigenous population.}
  \label{fig:flowchart}
\end{figure*}

To create two unique sets of domains for the indigenous and general populations, we conduct the process from step 1 to step 5 in two rounds. In the first round, we include keywords related to indigenous terms, such as \textit{Aboriginal and Torres Strait Islander}, to collect domains dedicated to indigenous services. In the second round, we exclude these terms to collect domains targeting the general public. After capturing these domains, we manually validate that they accurately match the respective target group. This process ensures the creation of two distinct and relevant domain sets. Data collected through this paper is accessible via IEEE DataPort \cite{dnsdataset}. Fig. \ref{fig:flowchart} illustrates the flowchart that outlines the steps to identify and categorize Australian government domains, focusing on services for the general population and indigenous communities.

\subsection{Retrieval of Authoritative Name Servers} \label{sec: Retrieval of Authoritative Name Server}
This section involves obtaining the authoritative name servers for the domains we collected. We develop a custom script based on command-line tools to obtain the DNS records, specifically focusing on NS records that indicate the authoritative DNS servers for each domain. We send queries to all 13 root servers \cite{RootServers} as of the 22nd of February 2023 \cite{IANA, Internic}.

Once we have obtained the name servers for the domains, our next objective in this stage is to establish two chains of dependencies---one for indigenous domains and another for general domains. This involves querying the name servers to acquire subsequent name servers, continuing this process until we reach the final authoritative name servers in the chain. This approach gives us a high chance of capturing the majority of the available NS records. Additionally, we leverage WHOIS to collect information about the provider company associated with each captured name server. The results obtained from this stage are utilized to visualize the DNS dependency graph in the next stage.

\subsection{Dependency Graph Construction} \label{sec: Dependency Graph Construction}

We construct two dependency graphs that illustrate the chain of dependencies between the DNS providers and the Australian government domains for both the general population and indigenous people. The are two types of relationships between domains and their name servers: direct and indirect dependencies. \textbf{Direct dependency} refers to a domain's immediate reliance on the name servers that are explicitly designated in its DNS configuration without considering any further dependencies of those name servers themselves.  
An indirect dependency, on the other hand, arises when the name servers directly associated with a domain further rely on other name servers or DNS providers for their own resolution. This creates a chain of dependencies involving intermediate providers.
This process results in a chain of dependencies from the domain/subdomain to the final authoritative name servers as well as the governmental websites and the final DNS providers, respectively. Fig. \ref{fig:dgraph_gen_exp} illustrates an example of how nodes are connected in the dependency graph. Optus is the DNS provider of \textit{gleneria.vic.gov.au.} Optus itself depends on Akamai as one of the hyperscaler-type providers. By visualizing the DNS dependency graph, we gain a clear and comprehensive understanding of how these entities are interconnected, allowing us to expose the impact of DNS dependencies on the indigenous and general domains.

\section{Results}
We present our findings on DNS dependencies and their geographical distribution for Australian government domains serving the general population and indigenous communities.
\subsection{Direct Dependency Analysis} \label{sec: Direct Dependency Analysis}

\begin{figure}[t]
  \centering  \includegraphics[width=\columnwidth,keepaspectratio]{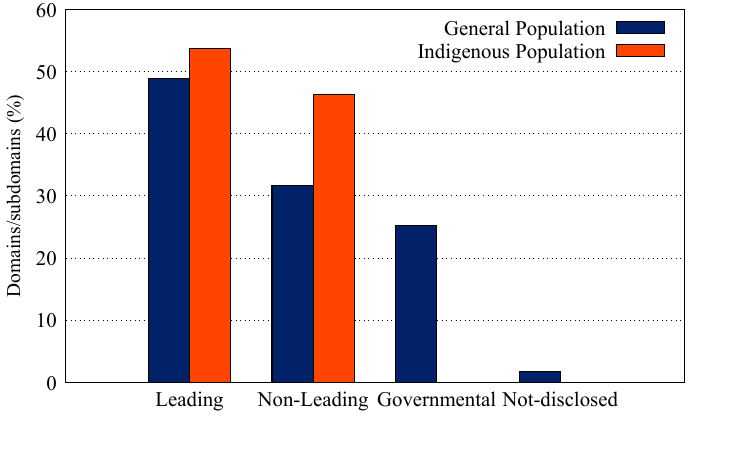}
  \caption{Dependency on various DNS provider types: general vs. indigenous domains.}
  \label{fig:category} \vspace{-2mm}
\end{figure}

We examine the immediate relationships between Australian government domains and their primary DNS providers. We analyze the distribution of these dependencies across different types of providers, including leading global companies, local Australian providers, and government-hosted services. Our findings reveal potential disparities in DNS provisioning between domains serving the general population and those targeted at indigenous communities.
In Fig. \ref{fig:category}, the dependency relationships between domains and DNS providers are illustrated, with providers classified as \textbf{leading}, \textbf{non-leading}, and \textbf{governmental} types. Leading DNS providers refer to the dominant and well-established companies in the industry, often associated with tech giants or hyperscalers. Examples of leading DNS providers mentioned in this paper include AWS Route 53, Microsoft Azure DNS, Cloudflare DNS, Google Cloud DNS, and Akamai Edge DNS. These providers have a significant market share and are known for their extensive infrastructure and global presence. Non-leading DNS providers, on the other hand, refer to smaller or lesser-known companies that offer DNS services. These providers may have a more limited market share compared to the leading providers and may operate on a regional or national scale. Examples of non-leading DNS providers in the Australian context are Telstra \cite{telstra}, Optus \cite{optus}, and Webcentral \cite{webcentral}. Governmental DNS providers, in this paper, are those that are operated and managed by the Australian government sector. In this paper, these providers are specific to the government sectors that are used to host DNS services for some government websites and domains. 

Table \ref{tab:summary} provides a summary of dependencies on third-party DNS providers for both general and indigenous domains.
The data is divided into two categories: general and indigenous populations, encompassing a total of 448 and 54 domains, respectively.  The analysis encloses various aspects of dependency, including the reliance on leading providers, non-leading providers (both domestic and international), as well as governmental providers. The analysis also examines two key aspects of DNS provider dependency: \textit{critical dependency} and \textit{diversified dependency}. Critical dependency refers to a domain's reliance on a single DNS provider for its name resolution services. When a domain critically depends on a single provider, it becomes vulnerable to any disruptions, outages, or security issues affecting that provider. In contrast, diversified dependency refers to a domain's reliance on multiple DNS providers, i.e., a site distributes its DNS infrastructure across different providers. This can enhance its resilience and availability, as a site is less likely to be impacted by issues affecting a single provider. It also allows for redundancy and fail-over mechanisms. Our results reveal a considerable amount of critical dependencies where just one DNS provider is used, with 92\% and 100\% for the general and indigenous domains, respectively. Specifically, 48.9\% of general domains and 53.7\% of indigenous domains rely on leading providers. Conversely, non-leading providers account for 31.3\% of dependency in the general population and 46.3\% in the indigenous populations. Moreover, the percentage of dependency on non-leading Australia-headquartered providers is almost doubled for indigenous people compared to the general domains. The dependency on non-leading non-Australian providers is relatively low for both groups at 8.5\% and 3.7\% for the general and the indigenous populations, respectively. Furthermore, the results reveal that 25.2\% of domains in the general population are dependent on intra-government providers, something we do not find for the indigenous group. Additionally, a critical dependency on non-leading Australian providers is observed in 20.8\% of general domains and 42.6\% of indigenous domains. These findings demonstrate a significant prevalence of critical dependencies on DNS providers in Australia, particularly within the indigenous population, where 100\% of domains are critically dependent on a single provider. For 2\% of the general domains, we were unable to identify the exact companies that act as DNS providers from the information in the DNS records or from the WHOIS (as information there was redacted).
Around 50\% of the domains serving both general and indigenous populations depend on leading DNS providers. This indicates that few providers handle a large portion of DNS services which could lead to potential risks of service outages during large-scale attacks. On the other hand, the other half of the indigenous services and 31.3\% of general services rely on non-leading DNS providers.

\begin{table}[t]
\centering
\caption{Direct and indirect dependencies on third-party DNS providers. Percentages reflect the proportion of the total number of domains that depend on the given provider type and do not sum to 100\% due to overlapping dependencies.}
\label{tab:summary}
\renewcommand{\arraystretch}{1.2}
\begin{tabular}{|p{30mm}|>{\centering\arraybackslash}p{12mm}|>{\centering\arraybackslash}p{11mm}|>{\centering\arraybackslash}p{14mm}|}
\hline
& \multicolumn{2}{c|}{\fontsize{7pt}{10pt}\selectfont\textbf{General(448)}} & \fontsize{7pt}{10pt}\selectfont\textbf{Indigenous(54)} \\
\hline
\rowcolor{gray!25}
Provider Type & Direct & Indirect & Direct \\
\hline
 Leading & 219(48.9\%) & 18(4\%) & 29(53.7\%) \\
\hline
 Non-Leading  & 140(31.3\%) & 12(2.7\%) & 25(46.3\%) \\
 \hline
 Non-Leading  (domestic) & 105(23.4\%) & 10(2.2\%) & 23(42.6\%) \\
 \hline
 Non-Leading (int'l.) & 38(8.5\%) & 4(0.9\%) & 2(3.7\%) \\
 \hline
 Intra-Governmental  & 113(25.2\%) & N/A & N/A \\
 \hline
 Not-disclosed & 8(1.8\%) & 3(0.7\%) & N/A \\
 \hline
\rowcolor{gray!25}
Critical Dependency & Direct & Indirect & Direct \\
\hline
 Any provider type  & 412(92\%) & 26(5.8\%) & 54(100\%) \\
 \hline
 Leading & 202(45.1\%) & 17(3.8\%) & 29(53.7\%) \\
 \hline
 Non-Leading & 122(27.2\%) & 9(2\%) & 25(46.3\%) \\
 \hline
 Non-Lead (domestic) & 93(20.8\%) & 7(1.6\%) & 23(42.6\%) \\
 \hline
 Non-Leading (int'l.) & 29(6.5\%)  & 2(0.4\%) & 2(3.7\%) \\
 \hline
 \rowcolor{gray!25}
Diverse Dependency & Direct & Indirect & Direct \\
\hline
 Any provider type  & 36(8\%) & 3(0.7\%) & N/A \\
 \hline
 Multiple Leading & 8(1.8\%) & N/A & N/A \\
 \hline
 Multiple Non-Leading & 2(0.4\%) & 2(0.4\%) & N/A \\
 \hline
 Multiple Gov sections & 6(1.3\%) & N/A & N/A \\
\hline
 Intra-Gov \& Leading & 1(0.2\%) & N/A & N/A \\
 \hline
 Intra-Gov \& Non-Leading & 18(4\%) & N/A & N/A \\
 \hline 
\end{tabular}
\end{table}

\begin{figure}[t]
  \centering
  \mathtoolsset{showonlyrefs=false}  \includegraphics[width=\columnwidth]{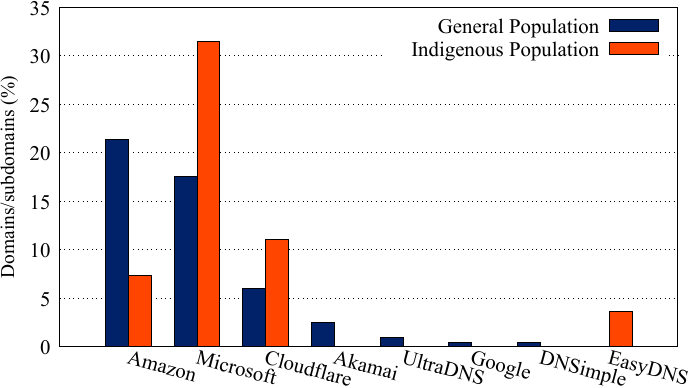}
  \caption{Dependency on leading DNS providers: general vs. indigenous domains.}
  \label{fig:leading} \vspace{-2mm}
\end{figure}

Fig. \ref{fig:leading} and Table \ref{tab:leading} specifically focus on the utilization of leading DNS providers by Australian governmental websites and show the percentages of domains that depend on specific leading providers. For instance, 21.4\% of general domains rely on Amazon DNS services, while only 7.4\% of indigenous domains do the same. The analysis also reveals the usage of other leading DNS providers, such as Microsoft, Cloudflare, Akamai, UltraDNS, Google, DNSimple, and EasyDNS, with varying proportions across the two population groups. That is, 31.5\% of indigenous websites use Microsoft DNS providers. For general domains, this is around 17\%. The rate of dependency on Cloudflare DNS providers for the indigenous population is 83\% more than the share of the general population. In contrast, the general population uses Akamai, UltraDNS (Neustar), Google, and DNSimple providers in less than 5\% of DNS services. However, indigenous websites do not use them. EasyDNS is only used by the indigenous-focused domains (5\%). The final row of the table indicates a relatively small number of general domains that depend on two DNS providers, namely Amazon and Microsoft. This configuration is not observed for the indigenous group.

\begin{figure}[t]
  \centering
  \mathtoolsset{showonlyrefs=false}  \includegraphics[width=\columnwidth]{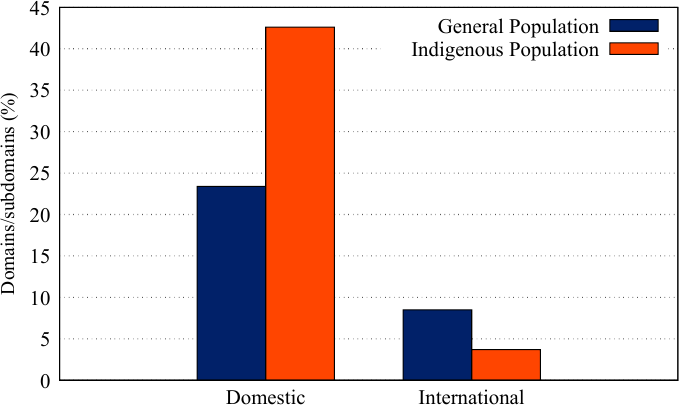}
  \caption{Dependency on non-leading DNS providers: general vs. indigenous domains.}
  \label{fig:non-leading} \vspace{-2mm}
\end{figure}

Similarly, the dependency on non-leading DNS providers utilized in Australia is shown in Fig. \ref{fig:non-leading}. Approximately 24\% of general domains rely on local DNS providers, whereas this proportion increases to over 40\% for indigenous domains. In contrast, indigenous websites mainly depend on less than 5\% of non-leading international DNS providers. This is close to 9\% for the general population. 

\begin{figure}[t]
  \centering
  \mathtoolsset{showonlyrefs=false}  \includegraphics[width=\columnwidth]{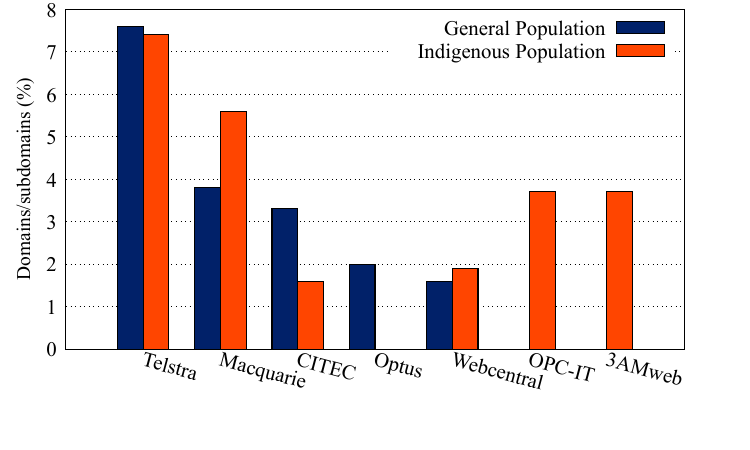}
  \caption{Dependency on most commonly used domestic DNS providers: general vs. indigenous populations.}
  \label{fig:domestic} \vspace{-1mm}
\end{figure}

Fig. \ref{fig:domestic} and Table \ref{tab:non-leading} focus on the most frequently used domestic DNS providers. These results indicate that a significant proportion of both the general and indigenous populations depend on non-leading DNS providers. Telstra emerges as the most commonly used DNS provider for both populations, followed by Macquarie Telecom and CITEC. Telstra, Australia's largest telecommunications company (by market share\cite{bib37_telstra}), has slightly less than 8\% of the share from both general and indigenous domains. The share of Macquarie Telecom is less than 4\% for the general population domains, and it is used by 6\% of indigenous websites. CITEC (Centre for Information Technology and Communication \cite{bib39_qldgov} \cite{bib38_citec}),  the Queensland Government's primary ICT services provider, has only a 4\% share of general domains. This number is only 2\% for the indigenous population. Webcentral has an equal share of less than 2\% from both general and indigenous domains. 4\% of indigenous DNS services are dependent on less-known Australian DNS providers such as OPC IT and ThreeAMWeb. General domains use Optus (which is a large subsidiary of a Singaporean telecommunications company). A significant outage of the Optus network occurred on 8 November 2023, affecting numerous customers and services across Australia \cite{optusoutage2024}. This incident shows the potential risks associated with relying on a single non-leading DNS provider. 
No indigenous domains were found to depend on non-Australian companies, while 14 general domains rely on the US-based company Verizon. These results show that there is a diverse range of DNS providers used by both populations, with some reliance on domestic providers and a limited reliance on non-Australian companies, and indigenous domains expose a higher reliance on non-leading providers.

\begin{table}[t]
 \caption{Leading DNS providers for the indigenous and general population
domains}
  \centering
  \begin{tabular}{lll}
    \toprule
    \textbf{Number of domains} & \textbf{General} & \textbf{Indigenous} \\
    \midrule
    Total & 219(48.9\%) & 29(53.7\%) \\
    Dependent on Amazon & 11(21.4\%) & 4(7.4\%) \\
    Dependent on Microsoft & 79(17.6\%)& 17(31.5)\%)  \\
    Dependent on Cloudflare & 27(6\%) & 6(11.1\%) \\
    Dependent on Akamai & 11(2.5\%) & 0 \\
    Dependent on UltraDNS & 4(0.9\%) & 0  \\
    Dependent on Google & 1(0.2\%) & 0  \\
    Dependent on DNSimple & 1(0.2\%) & 0  \\
    Dependent on EasyDNS & 0 & 2(3.7\%)  \\
    Dependent on both Amazon and Microsoft & 6(1.3\%) & 0 \\
    \bottomrule
  \end{tabular}
  \label{tab:leading}
\end{table}

\begin{figure}[t]
  \centering
  \mathtoolsset{showonlyrefs=false}  \includegraphics[width=0.9\columnwidth]{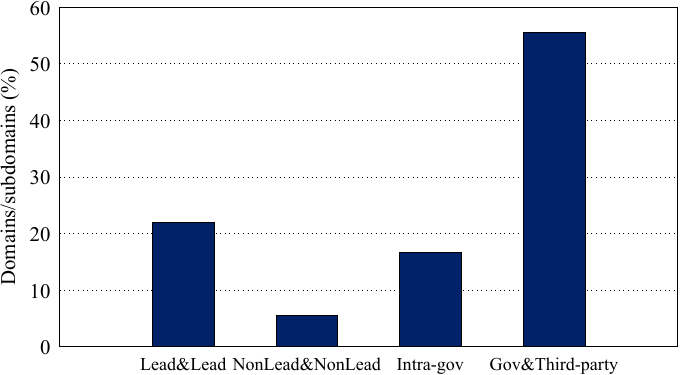}
  \caption{Diversified dependencies in general domains.}
  \label{fig:multi} \vspace{-3mm}
\end{figure}

In Fig. \ref{fig:multi}, we illustrate how domains are distributed among the general population with multiple providers. Notably, none of the indigenous domains are dependent on multiple DNS providers. As shown in this figure, 20\% of domains have two distinct leading DNS providers (Amazon and Microsoft), and less than 5\% have two different non-leading providers. Additionally, less than 20\% of domains have intra-government DNS providers that point to different government sectors. Over half of general domains use both a government-hosted provider and a third-party DNS provider.

\begin{table}[t]
  \caption{Direct/indirect dependencies, general domains.}
  \centering
  \begin{tabular}{lll}ule
    \textbf{No. of domains} & \textbf{Direct} & \textbf{Indirect} \\
    \midrule
    5 & Not-disclosed & Leading \\ 
    1 & Intra-gov & Leading \\ 
    11 & Intra-gov & Non-leading (domestic\&international) \\ 
    11 & Intra-gov & Non-leading (domestic) \\ 
    2 & Intra-gov & Non-leading (international) \\ 
    12 & Non-leading & Leading \\ 
    2 & Non-leading & Non-leading (international) \\
    \bottomrule
  \end{tabular}
  \label{tab:indirect}\vspace{-2mm}
\end{table}

\subsection{Indirect Dependencies} \label{sec: Indirect Dependency}
Indirect dependencies represent a more complex web of relationships in DNS provisioning. We investigate these to disclose hidden vulnerabilities and centralization patterns that may not be evident from direct dependencies alone. We explore how indirect dependencies occur in general domains (Table \ref{tab:indirect}), focusing on the relationship between a domain's immediate DNS provider type (namely, leading and non-leading provider, with either Australian headquarters or non-Australian headquarters) and its reliance on the indirect DNS provider type, and discuss the implications for service reliability and security. In our observation, there is no evidence of indirect dependencies for government domains dedicated to indigenous people.
Our examination reveals a dependence on major companies such as Amazon and Akamai.
Interestingly, one domain employing government-hosted DNS services also has an indirect dependency on Amazon.
While certain domains directly depend on intra-government services, their indirect dependencies often lie with non-leading, both domestic and international, providers such as Telstra, Optus, and CITEC. This diversity in indirect providers could offer redundancy and reduce single points of failure while also introducing complexity and potential exposure to vulnerabilities.
A significant observation is the indirect dependency of 12 domains on leading providers despite their direct reliance on non-leading ones. This pattern underscores the pervasive influence of hyperscalers such as Amazon and Akamai in the DNS infrastructure. The contrast between direct and indirect dependencies reveals a critical aspect of DNS infrastructure management. For general population domains, this leads to centralization around a few hyperscaler entities.

\begin{table}[tb]
  \caption{Non-leading DNS providers.}
  \centering
  \begin{tabularx}{\columnwidth}{lXXX}
    \toprule
    \textbf{Number of domains} & \textbf{General} & \textbf{Indigenous} \\
    \midrule
    Total & 142(31.7\%) & 25(46.3\%) \\
    Dependent on Telstra & 34(7.6\%) & 4(7.4\%) \\
    Dependent on Macquarie Telecom & 17(3.8\%)& 3(5.6\%)  \\
    Dependent on CITEC & 15(3.3\%) & 1(1.6\%) \\
    Dependent on Verizon  & 14(3.1\%) & 0 \\
    Dependent on Optus  & 9(2\%) & 0 \\
    Dependent on Webcentral  & 7(1.6\%) & 11(1.9\%) \\
    Dependent on NEC  & 7(1.6\%) & 0 \\
    Dependent on OPC IT  & 0 & 2(3.7\%) \\
    Dependent on Three AM Web  & 0 & 2(3.7\%) \\
    \bottomrule
  \end{tabularx}
  \label{tab:non-leading}
\end{table}

\subsection{Geographical Distribution of DNS Servers}
The physical location of DNS servers plays a crucial role in the performance and resilience of online services. By examining the geographical distribution of IP addresses associated with DNS servers, especially those serving specific domains such as Australian government domains dedicated to indigenous peoples, we can disclose patterns of reliance on third-party providers. 
 We map the geographical distribution of DNS servers associated with Australian government domains and analyze how server locations differ for domains serving the general population versus those focused on indigenous communities. For this, we leveraged the \textbf{IP2Location} database. In addition to the geographical location, IP2Location also provides information about the usage type of IP addresses. The usage types relevant to our analysis are as follows: 1) ISP (Internet Service Provider): IP addresses associated with Internet service providers 2) DCH (Data Center/Web Hosting/Transit): IP addresses belonging to data centers, web hosting companies, or transit providers (the latter category is less likely to contain authoritative name servers) 3) CDN (Content Delivery Network) 4) GOV (Government): IP addresses assigned to government entities, implying that the DNS server is operated by a government organization.
 
 A problem that is common to geolocation databases is that IP addresses, especially of hyperscalers, may be attributed to the wrong country, in addition to IP allocations also ever-changing (which may change the usage type). Given our small sample size, our results must be taken with a grain of salt. For example, the geographical distribution of the DNS servers associated with Australian government domains for indigenous peoples spans across five countries: Australia, Canada, Hong Kong, Japan, and the United States. The Australian setting is much more plausible than that of Hong Kong or Japan, which may be such cases of wrong attributions. However, it would still imply the use of powerful providers and hyperscalers. In Australia, the presence of ISP/DCH usage types indicates a mix of local Internet service provisioning and data center hosting. The United States is also a plausible hosting location: it hosts the highest number of DNS servers, indicating a heavy reliance on US-based infrastructure. However, it is still possible that these servers are physically located in Australia. An international dependency may well be more vulnerable, both from a technical and operational point of view.

\begin{figure}[tb]
  \centering
  \mathtoolsset{showonlyrefs=false}  \includegraphics[width=\columnwidth]{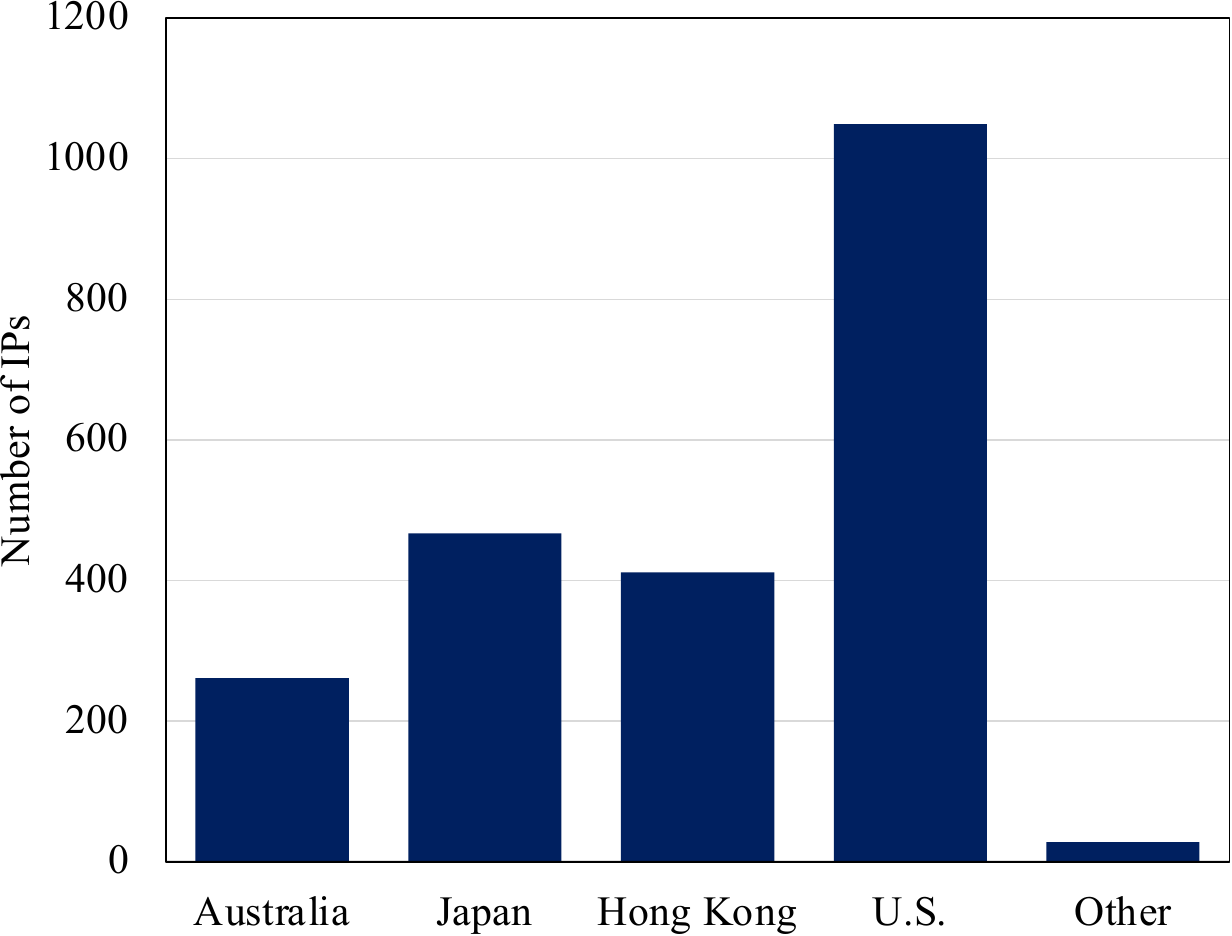}
  \caption{Geographical location of general domains' NS.}
  \label{fig:geo_direct_gen}\vspace{-2mm}
\end{figure}

\begin{figure}[tb]
  \centering
  \mathtoolsset{showonlyrefs=false}  \includegraphics[width=\columnwidth]{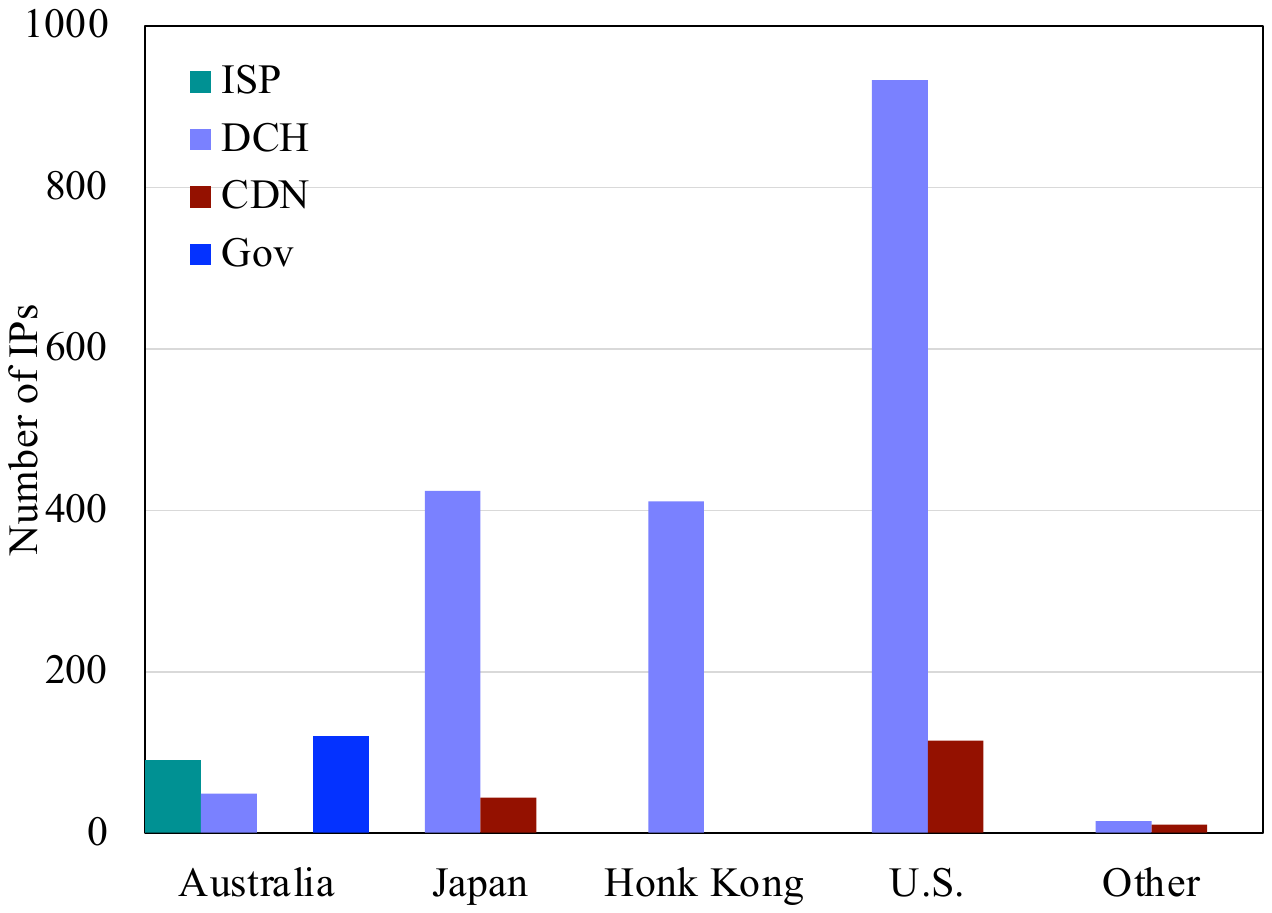}
  \caption{Usage type for general domains' NS.}
  \label{fig:usagetype_direct_gen}\vspace{-3mm}
\end{figure}

\begin{figure}[tb]
  \centering
  \mathtoolsset{showonlyrefs=false}  \includegraphics[width=\columnwidth]{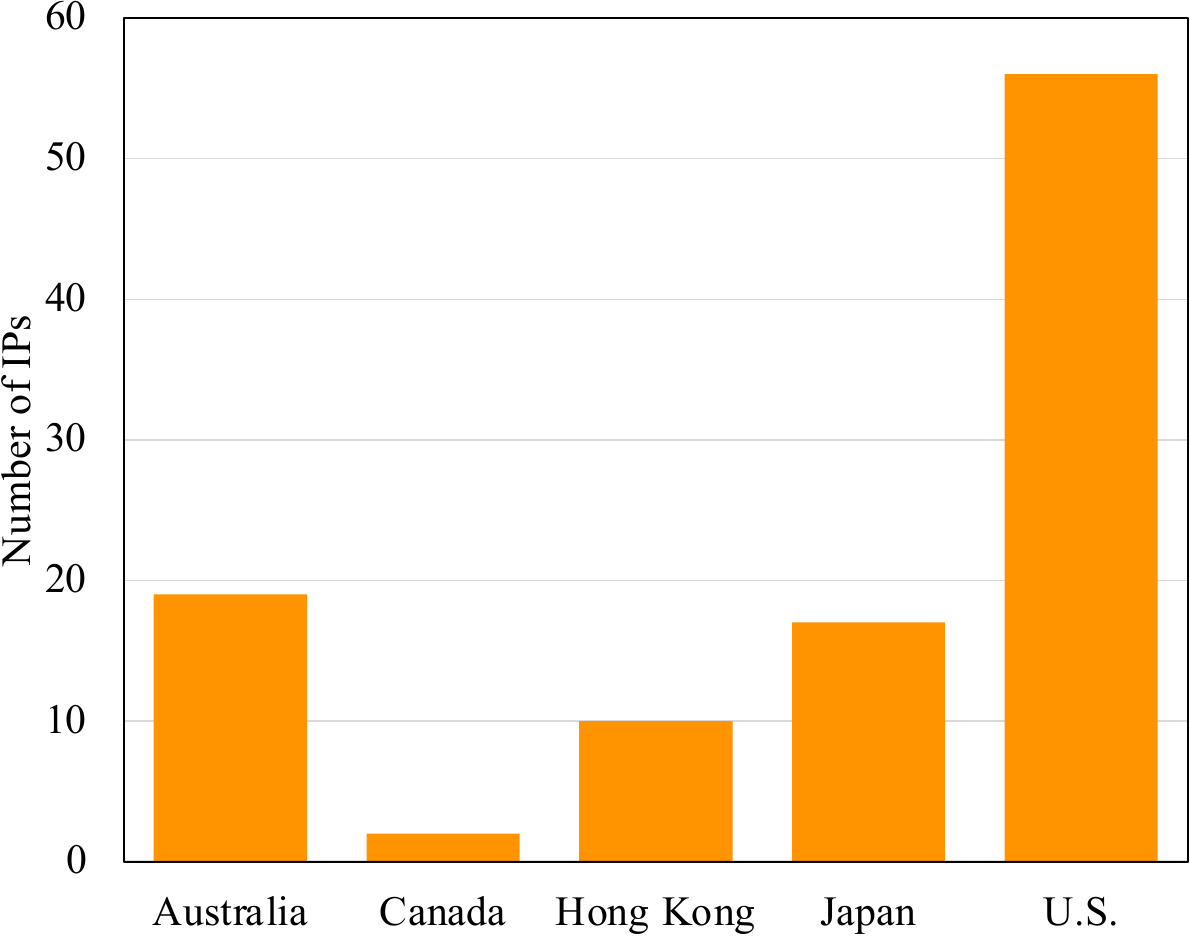}
  \caption{Geographical location of indigenous domains' NS.}
  \label{fig:geo_direct_indi}\vspace{-2mm}
\end{figure}

\begin{figure}[t]
  \centering
  \mathtoolsset{showonlyrefs=false}  \includegraphics[width=\columnwidth]{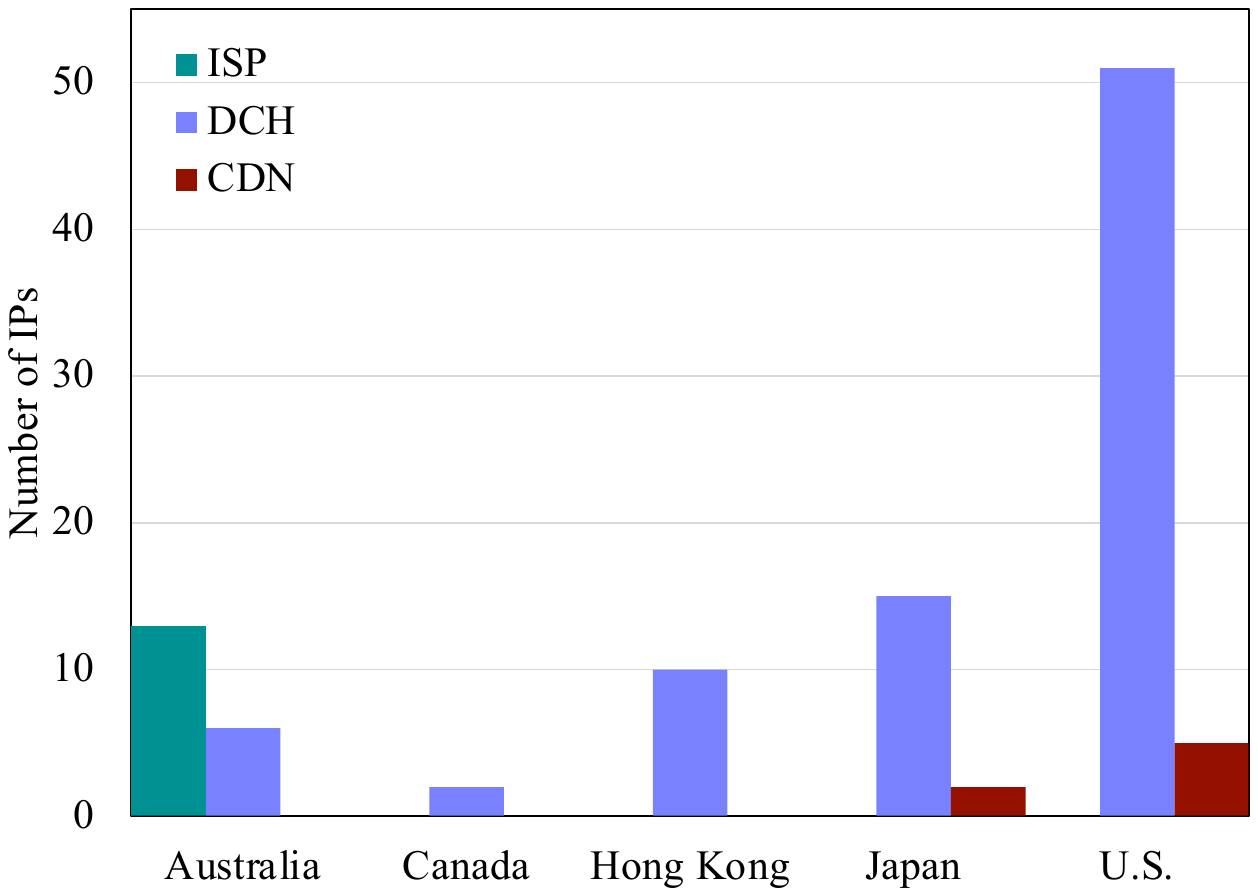}
  \caption{Usage type for indigenous domains' NS.}
  \label{fig:usagetype_direct_indi}\vspace{-3mm}
\end{figure}

Both domain groups have DNS servers primarily located in Australia and the United States, as shown in Fig. \ref{fig:geo_direct_gen}, \ref{fig:usagetype_direct_gen}, \ref{fig:geo_direct_indi}, and \ref{fig:usagetype_direct_indi}.
The breakdown into usage types in Fig. \ref{fig:usagetype_direct_gen} indicates an overall diversified approach to DNS service provisioning for the general domains. The presence of ISPs suggests localized service delivery, while CDNs (Content Delivery Networks) highlight efforts to optimize content distribution and performance. A small proportion is classified under ``Gov'', indicating government-managed servers, which may signify an intention for self-reliance and control over DNS infrastructure. This hints at a balance between globalized service provision and localized service delivery and at strategic approaches to deployment and performance optimization.
The presence of government-managed servers is also noteworthy, especially as they are not present in the indigenous-focused infrastructure. The DNS infrastructure for the general population seems to have a greater emphasis on diversity and global distribution compared to that of the indigenous peoples. The inclusion of government servers for the general population suggests an element of direct governmental control or self-reliance. For the indigenous-focused domains, the emphasis appears to be on a mix of local and international service provisioning with a significant dependency on DCH and CDN providers.

\section{Discussion} \label{sec: Discussion}
Our analysis of DNS dependencies for Australian government domains reveals implications for the digital divide, geographic dependencies, and security vulnerabilities. We discuss these findings in the context of our three main research questions.

\subsection{Digital Divide}
Our results provide evidence of a digital divide in DNS provisioning between general and indigenous domains. The main difference is the critical dependency on single providers: 100\% of indigenous domains rely on a single DNS provider, compared to 92\% of general domains. This higher dependency makes indigenous domains more vulnerable to service disruptions and potential attacks. Furthermore, indigenous domains show a greater reliance on non-leading DNS providers (46.3\%) compared to general domains (31.3\%). While this might reflect a preference for local services, it also exposes these domains to potential reliability issues, especially given that some of these providers may have limited resources for security and maintenance. Surprisingly, none of the indigenous domains utilize government-operated DNS infrastructure. This absence requires further investigation to understand its causes and implications for security and reliability.

\subsection{Geographic Dependencies}
The geographical distribution of DNS servers reveals another aspect of DNS provisioning, although we caution that geolocation information can be misleading, and hence our results must be viewed in that light. Nevertheless, some trends are evident. General population domains benefit from a broader global spread. In contrast, indigenous domains show a higher concentration of servers in fewer locations. The limited geographic diversity for indigenous domains could lead to increasing vulnerability to regional outages, imposing potential latency issues for users in different regions and greater exposure to geographically targeted attacks. The preference for domestic DNS providers by over 40\% of indigenous domains, while potentially driven by a desire for geographically closer services, may increase the vulnerability to region-specific disruptions or attacks.

\subsection{Security Analysis}
For our analysis of how various forms of dependencies lead to a site being more or less vulnerable, we define three different attacker models.  We categorize potential attackers into three representative groups of varying strengths: lone wolves (including individual attackers) \cite{kessler2015lone, kessler2016lone}, the organized fringe attackers (such as hacktivists) \cite{jing2003areca, girtler2003fringe, davies2018characterizing, taylor2005hackers}, and global actors (e.g., nation states) \cite{sheldon2012state, pawlicka2020cyberspace}. Each group has different resources, intentions, and targets, ranging from limited-resource attackers targeting specific populations to well-funded, expert attackers operating on a global scale.

We call our first attacker ``lone wolf---weak but targeted''. This attacker does not have many resources (hardware, network access, deep technical knowledge, finances) at their disposal. However, they act out of a strong, personal motivation, and they only target particular organizations. An example of such strong motivation would be hate of a particular societal minority (\eg racist sentiments, resentment of the LGBTQ+ community, \etc). Their likely target is organizations representing or aiding such minorities, and their tools are limited to what limited financial means can afford---for example, paying an underground service a small amount of money to stage a limited DDoS attack. In the context of our results, this attacker is more concerned with the indigenous domains. The lack of DNS service diversity makes indigenous domains particularly vulnerable. Even with limited resources, a lone wolf attacker could potentially disrupt services by focusing on the single provider used by a domain.

We call our second attacker ``organized fringe---limited and targeted''. This attacker is typically a group that can raise some relevant resources (financial contributions by members, donations) and gain access to relevant knowledge that can be used in an attack (\eg members with a deep technical background). The motivation is broader than in the previous attacker model and represents, for instance, political fringe movements: anti-democratic forces that also target minority groups or direct resentment at them in a bid to win more support. The likely targets are the same as before as well as various government services. This attacker can invest substantially more resources---a more powerful DDoS attack is a possibility, as is a takeover of the IT services of some smaller companies with less stringent security. This attacker type is of concern for both domain groups, but still more so for the indigenous domains. The concentration of indigenous domains on Australian providers makes them particularly attractive targets.

We call our third attacker a ``global actor---resourceful and impact-oriented''. This attacker has means at their disposal that one would usually associate with organized crime or, at worst, even rogue nation-states. They can make substantial investments into sustained attacks and have access to skilled hackers. The motivation here may be much broader than before: the attacker may be less interested in hurting smaller organizations and more focused on causing widespread damage that has economic ramifications and will draw attention from politics, often at a global level. Damage to services aiding minorities may be a welcome side effect, but the targets would more commonly be large, professional enterprises that offer cloud services---in the more extreme cases, even global hyperscalers, such as the DDoS attack on Amzon's DNS service in October 2019 \cite{williams2019bezos}.  While these sophisticated attackers pose a threat to all domains, the lack of diversity and redundancy in indigenous domain DNS infrastructure makes them particularly vulnerable to widespread, high-impact attacks.

To mitigate these risks, we recommend the following: (1) Implementing multi-provider setups for all domains, especially indigenous ones. (2) Diversifying the geographic locations of DNS servers. (3) Incorporating government-operated DNS services as an additional layer of resilience.
These measures would enhance the security against at least the two less powerful forms of attackers and improve overall service reliability.

\section{Summary} \label{sec: Summary}
This paper investigated the impact of Internet centralization on DNS provisioning and its security implications, particularly focusing on the indigenous and general populations in Australia. The study categorized DNS providers into leading (hyper-scaler, US-headquartered companies), non-leading (smaller Australian-headquartered or non-Australian companies), and Australian government-hosted providers and constructed dependency graphs to demonstrate direct and indirect dependencies of Australian government domains on these providers.

Our main findings included as follows: (1) Digital Divide: There is a significant difference in the DNS service providers used for indigenous and general population domains, which imposes potential disparities in service quality and reliability. Indigenous domains revealed a higher dependency on non-leading DNS providers compared to general population domains, indicating a digital divide in DNS provisioning. (2) Geographic Dependencies: The geographical distribution of DNS servers revealed the concentration of servers in fewer locations for indigenous domains, which could lead to increased vulnerability to regional outages and potential latency issues. The DNS servers for general population domains benefit from a broader global spread. (3) Security Problems: The study demonstrated vulnerabilities of each domain group to different attacker models. Indigenous domains are more vulnerable due to their critical dependency on single providers and limited geographic diversity. The analysis categorized potential attackers into lone wolves, organized fringe attackers, and global actors, each posing different levels of threat to the DNS infrastructure.

The paper implied that addressing these disparities and vulnerabilities requires implementing multi-provider setups, diversifying geographic server locations, and incorporating government-operated DNS services to enhance security and service availability, which could be the future research effort on DNS dependency.

\newpage
\bibliography{references.bib}{}

\bibliographystyle{ieeetr}

\vspace{-15 mm}
\begin{IEEEbiography}
    [{\includegraphics[scale = 0.8]{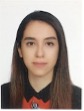}}]{Niousha Nazemi}

is currently a PhD student in the School of Computer Science, University of Sydney, Australia. Her main research interests include Internet centralization, DNS measurement, and security of federated learning.
\end{IEEEbiography}

\vspace{-13 mm}
\begin{IEEEbiography}
    [{\includegraphics[scale=0.065]{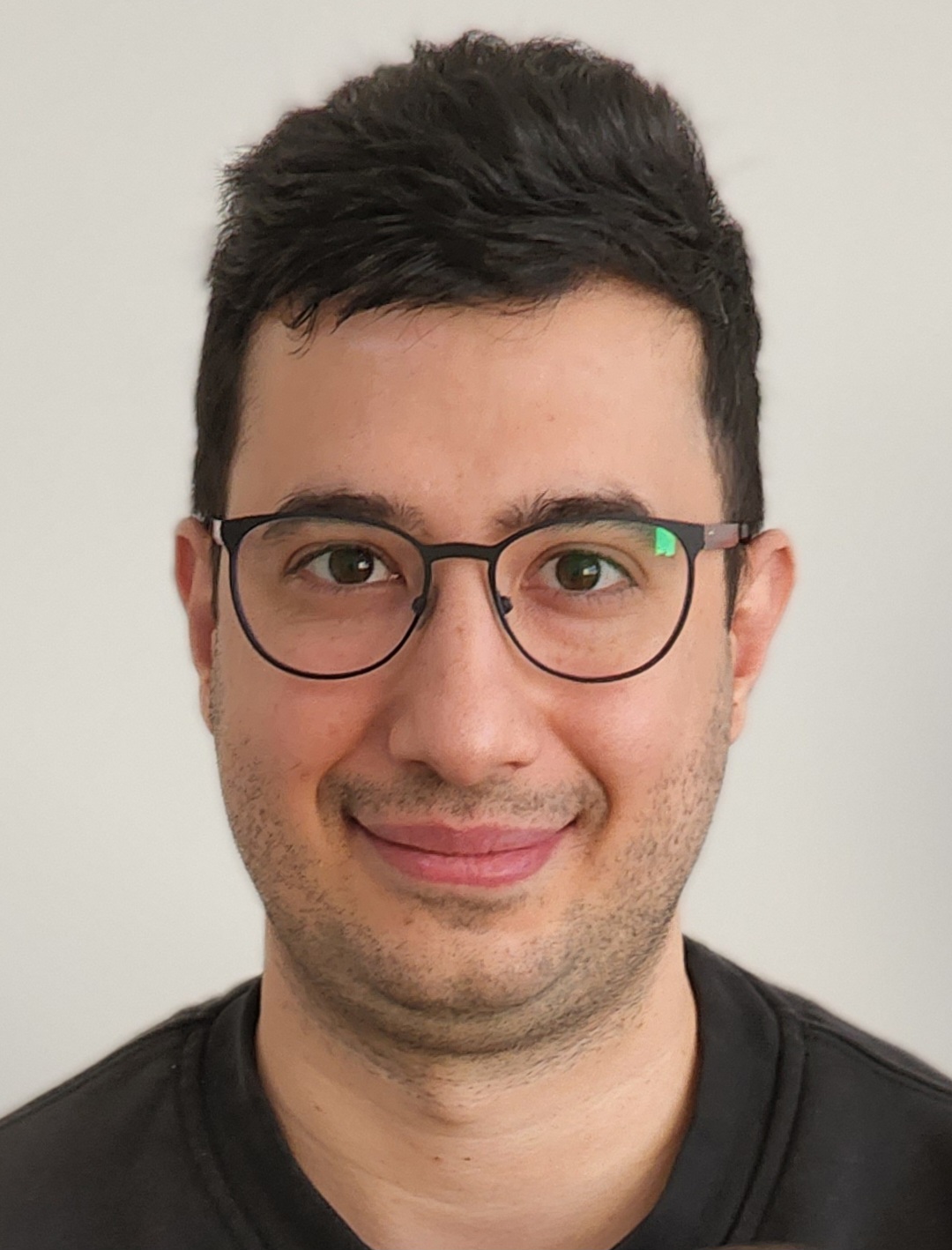}}]{Omid Tavallaie}
    
is a Postdoctoral Researcher in Computer Networking at the Department of Engineering Science, University of Oxford. Prior to joining the University of Oxford, he was a Research Associate and Lecturer at the School of Computer Science, The University of Sydney, and a member of the Centre for Distributed and High-Performance Computing (Australia). He holds a PhD in Computer Science from the University of Sydney in 2021.
\end{IEEEbiography}

\vspace{-13 mm}
\begin{IEEEbiography}
    [{\includegraphics[scale=0.13]{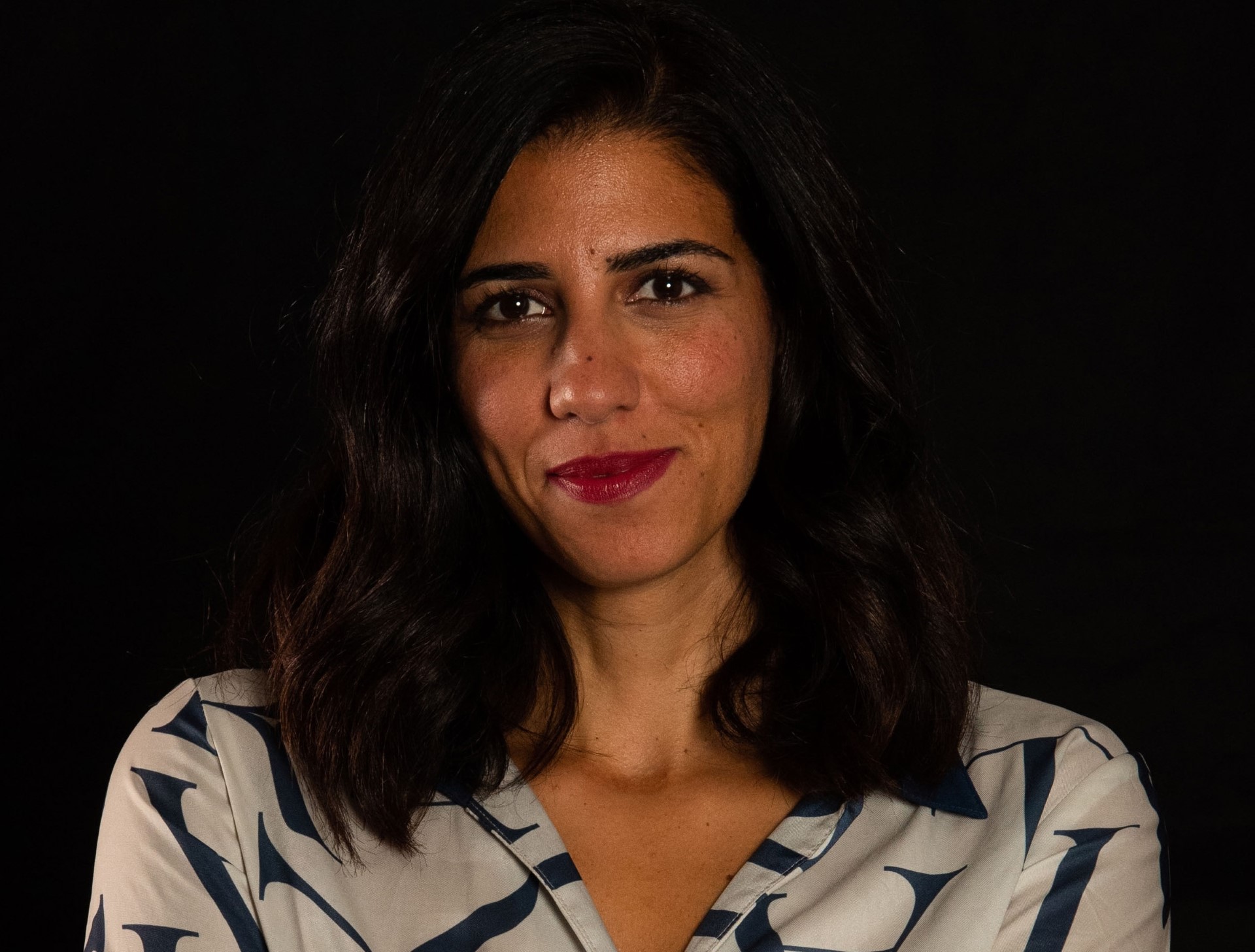}}]{Anna Maria Mandalari}
    
works as Assistant Professor at University College London (UCL). She is affiliated with the Electronic \& Electrical Engineering Department. She is member of the Italian Technical Secretariat of the Committee for strategies on the use of AI. She is Honorary Research Fellow at the Institute for Security Science and Technology at Imperial College London and expert fellow of the UK SPRITE+ Hub. She obtained her PhD within the framework of the METRICS project, which is part of the Marie Skłodowska-Curie action, intended for excellent researchers, affiliated with the Carlos III University of Madrid. Her research interests are Internet of Things (IoT), privacy, security, networking, Internet measurement techniques, and AI. She studies privacy implications and information exposure from IoT devices. She works on the problem of modelling, designing, and evaluating adaptation strategies based on Internet measurements techniques. In addition to her research, Anna gives invited talks all around the world to promote research and create awareness on security, privacy, and ethical AI. Most of her research experiences have also significantly contributed to several EU-funded research projects and have had a significant influence on media and policymaking.
\end{IEEEbiography}

\vspace{-13 mm}
\begin{IEEEbiography}
    [{\includegraphics[scale=0.1]{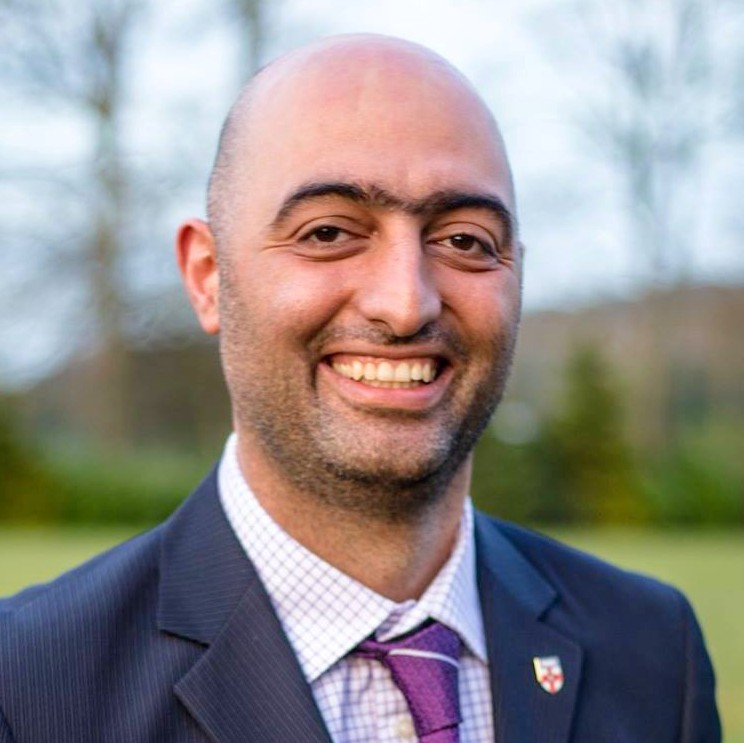}}]{Hamed Haddadi}
    
(Member, IEEE) is a Professor in Human-Centred Systems at the Department of Computing at Imperial College London. He is part of Imperial-X (I-X) where he leads the Privacy and Security Research Area. He also serves as a Security Science Fellow of the Institute for Security Science and Technology. In his industrial role, he is the Chief Scientist at Brave. His research interests are in User-Centred Systems, IoT, Applied Machine Learning, Privacy, and Human-Data Interaction.
\end{IEEEbiography}

\begin{IEEEbiography}
    [{\includegraphics[scale=0.45]{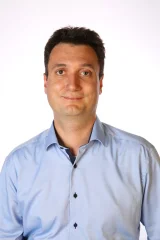}}]{Ralph Holz}
    
is a Full Professor at the University of Münster in Germany, and co-appointed at the University of Twente in the Netherlands. His research interests revolve around empirical security analysis, with a focus on Internet security.
\end{IEEEbiography}

\vspace{-13 mm}
\begin{IEEEbiography}
    [{\includegraphics[scale=1.35]{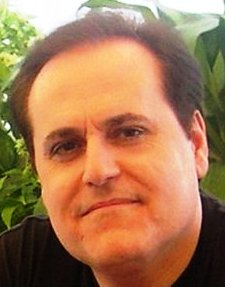}}]{Albert Y. Zomaya}
    
(Fellow, IEEE) is the Peter Nicol Russell Chair Professor of Computer Science in the School of Computer Science, Sydney University, and the Director of the Centre for Distributed and High-Performance Computing. Professor Zomaya has published over 700 scientific papers and articles and is the author, co-author, or editor of more than 30 books. He is a decorated scholar with numerous accolades, including Fellowships of the IEEE, AAAS, and the IET. Also, he is a Fellow of the Australian Academy of Science, a Fellow of the Royal Society of New South Wales, a Foreign Member of Academia Europaea, and a Member of the European Academy of Sciences and Arts. His research interests include parallel and distributed computing, networking, and complex systems.
\end{IEEEbiography}
 \end{document}